\documentclass[11pt]{article}
\newcommand{\al}{\alpha}

\newcommand{\ice}[1]{\relax}
\newcommand{\be}{\begin{equation}}
\newcommand{\ee}{\end{equation}}
\newcommand{\ba}{\begin{eqnarray}}
\newcommand{\ea}{\end{eqnarray}}
\newcommand{\nn}{\nonumber}
\newcommand{\MSsch}{{\overline{\rm MS}}}
\newcommand{\GeV}{{~\rm GeV}}

\begin{document}
\begin{flushright}
INR-TH/03-3\\
January 2003
\end{flushright}
\begin{center}
{\Large \bf Determination of the numerical value for the
strong coupling constant from tau
decays in perturbative QCD\footnote{
Talk given at 15th International Seminar ``Quarks-2002'',
Novgorod the Great, Russia, June 1-7, 2002}
}

\vskip 1.2cm
{\large \bf A.A.~Pivovarov}\\[2mm]
Institute for Nuclear Research of the
  Russian Academy of Sciences,

Moscow 117312
\vskip 1cm

{\bf Abstract}
\end{center}
\vskip -0.3cm
The determination of the numerical value for the
strong coupling constant from tau
decays in perturbative QCD is briefly described.
Main emphasis is made on the use of the
renormalization scheme invariance of the theory for reducing
uncertainties
related to truncation of the series in the strong coupling constant.
The analysis of the convergence of the series is presented and
the possibility of the asymptotic growth at the NNNLO is discussed.

\vskip 1cm
In this talk I briefly describe one aspect of $\tau$ physics --
the use of data on $\tau$-lepton hadronic decays for extracting
a numerical value of the strong coupling constant that is mainly
based on the recent analysis presented in ref.~\cite{Korner:2000xk}.
The accuracy of experimental data is steadily improving
that provides a good opportunity for high precision tests
of the theoretical description~\cite{exp1,exp2,PDG}.
The underlying theory of strong
interactions -- QCD~\cite{qcd} -- describes the observables related to the
physics of $\tau$-lepton hadronic decays within perturbation theory in
the strong coupling constant, i.e. as a series expansion.
The key question of theoretical description is a pattern of
convergence of the related series. On general basis, it is believed
that the series is asymptotic, however, in some concrete applications
it can appear as a convergent series.
Thus, establishing the accuracy with which the series, truncated in a
given manner, represents the ``full'' result of
perturbation theory calculation is a
central question of using pQCD in low-energy phenomenology.
The $\tau$-lepton hadronic decays is a good place to investigate this
problem since the theoretical description of the process is very clean.
Indeed, the basic quantity is the two-point correlator of
hadronic currents in the Euclidean domain which has been
calculated with a very high
degree of accuracy within perturbation theory~\cite{eek20,eek21,eek2c}
(for a review see~\cite{phys_report}).
Nonperturbative corrections to the correlator are known to
be small and under control within the operator product
expansion~\cite{SVZ}. For comparison of experimental data with
theoretical calculations one can choose inclusive observables
that makes pQCD directly applicable to the description of the
system~\cite{SchTra84,Bra00,NarPic88,Pivtau,BraNarPic92,DP}.
These features explain the interest attracted by the $\tau$-lepton
hadronic decays for extracting numerical values of the
parameters of the standard model at lower energies.
These values can then be compared with the results of high energy
experiments that provides a powerful consistency check of QCD
(and the standard model as a whole)
for energies from $M_\tau=1.777\GeV$ till
$M_{Z}=91.187\GeV$ (e.g.~\cite{wilczek}).

I discuss in some detail the procedure of extracting numerical values of $\al_s$
from $\tau$-data in perturbation theory.
Since the numerical value of the expansion
parameter $\al_s$ is not small at the $M_\tau$ scale
the contribution of higher order terms in the
perturbation theory series can be significant.
There are arguments that the accuracy
of finite-order perturbation theory is already close
to its asymptotic limit~\cite{tau12}.
Therefore a kind of resummation is necessary.
A powerful technique is $1/N_F$ expansion (e.g.~\cite{qed})
that leads to renormalon-type
resummation~\cite{Mueller:vh,renZakh,bigi2,beall,katmax,cvetic}
that can be formulated as an integral of the running coupling constant
in momentum space. This technique is often used in the analysis
based on Dyson-Schwinger
equations~\cite{Krasnikov:1982gp,Alkofer:2000wg}.
The result of the resummation under the integration sign
depends strongly on the interpretation
of higher order terms, e.g~\cite{renRS,Faleev:pg,Epele:2002ce}.
The resummation of contributions related to the running of
the coupling constant in QCD was used as
a basis for the tau lepton decays analysis in contour improved
perturbation theory~\cite{Pivtau,resana}.

The central point to be discussed in the talk
is an explicit use of renormalization group invariance
of the theory in order to analyze the
$\tau$-lepton decay rate in the most optimal way within perturbation theory.
Renormalization group invariance is a fundamental property
of perturbation theory in quantum field theory
which is (formally) related to the freedom
in defining the subtraction procedure~\cite{RG}.
Its explicit compliance may make a numerical analysis
more reliable and stable.
Note that it can be viewed in a broader context as an optimal choice
of the starting point for the perturbative expansion, i.e.
the way of splitting the whole Hamiltonian into two parts:
the leading one that is
considered to be large and
``perturbation'' that is considered to be small and can be taken
within perturbation theory. The influence of the proper choice of the
initial approximation on the accuracy of the analysis
can be clearly seen in model examples~\cite{Penin:1996bj} or in
the framework of effective theories~\cite{vol,hoang}.
In QCD the renormalization group invariance allows one to formally
perform the numerical analysis in any renormalization scheme
because all schemes are connected by a renormalization group
transformation~\cite{stevenson,Brodsky:1982gc}.
In the finite-order perturbation theory approach
this equivalence is only approximate
and broken by higher order terms
which
introduce numerical differences into the results
obtained in different renormalization schemes
for the same quantities.
Generally there many ways of using perturbation theory
calculations.
One is to find direct relations between physical observables
which are renormalization group invariant.
Then perturbation theory calculations are just
a purely intermediate step for finding
formal relations between observables (see, e.g.
\cite{tau12,prl}) and no numerical analysis
for renormalization scheme noninvariant quantities is performed.
Indeed, let the perturbation theory expressions
for two observables ${\cal O}_{1,2}$
in a given scheme have the form
\be
\label{twoO}
{\cal O}_1=\al_s+r_1\al_s^2+O(\al_s^3), \quad
{\cal O}_2=\al_s+r_2\al_s^2+O(\al_s^3) \, .
\ee
Then the perturbation theory
relation between observables ${\cal O}_{1,2}$ reads
\be
\label{reltwoO}
{\cal O}_2={\cal O}_1+(r_2-r_1){\cal O}_1^2+O({\cal O}_1^3)
\ee
and is scheme-independent.
The difference $r_2-r_1$ takes the same value for calculations in any scheme.
Another way of using the result of perturbation theory
calculations is to extract numerical values for
renormalization scheme noninvariant
quantities (as the coupling constant in a fixed scheme)
and then use it for predictions of different observables.
The truncation of the perturbation theory series
leads to numerical
violations of renormalization scheme invariance
and plays an essential role.
In our simple example this means that the relations in eq.~(\ref{twoO})
are treated as quadratic functions of $\al_s$ in some fixed scheme and the
accuracy of extraction of the coupling constant value (and prediction
of other observables) depends drastically on the scheme
used, i.e. on the numerical values of the coefficients $r_{1,2}$.
This happens because the accuracy is estimated in a heuristic way by
considering the apparent convergence or the magnitude of the last term.
There is an analogy with gauge invariance (or any other symmetry):
one can either work with invariant quantities or fix the gauge and
compute in terms of noninvariant quantities but the final result for
physical observables (in a gauge invariant sector) will be the same.
In practical application it is realized, for instance, as calculation
in covariant or external field gauges.

The concrete example of this talk is the
extraction of a numerical value
for the coupling constant which is not an immediate physical quantity.
By convention the reference value of the coupling constant
that is used to compare between
different experiments is fixed to be the $\MSsch$-scheme one.
However,
this does not necessarily
mean that for its extraction from a given experiment
the numerical analysis
should be performed in the $\MSsch$-scheme.
It can be numerically accurate to
analyze the system in its internal scheme and after finding numerical
values for the internal parameters translate them into
the $\MSsch$-scheme using
renormalization scheme transformation.
This program is based on explicit renormalization scheme covariance of
the theory (as calculations
in different gauges are only possible
(i.e. meaningful) for
the gauge invariant sector of gauge theories).
Note that the analogy with gauge is not quite complete:
gauge invariance is exact order by order in perturbation theory while
the renormalization scheme
invariance is only exact for a 'full' quantity while
for truncated series it is only accurate precise with accuracy the order of
the value of the first omitted term.
Note also that gauge invariant quantities could not always be put
into the same order of perturbation series.
The approach used below, in practice, means that any scheme is allowed for the
analysis of a particular quantity while the transition between
the schemes is considered to be exact.
Therefore numerical values
obtained in the $\MSsch$-scheme
directly and through renormalization group transformations can differ.
In fact, in the spirit of the perturbation theory
the internal scheme results are most reliable
physically and are more stable numerically
than the results of the standard analysis in the $\MSsch$-scheme.
The numerical values for the reference $\MSsch$-scheme parameters are
obtained by a renormalization group
transformation from the numerical values found in the internal schemes.
Renormalization group transformation
is a quite formal operation and can be controlled
numerically. For instance, the
change of scale is a renormalization group transformation.

We consider a simplest example that exhibits relevant
features --the normalized
$\tau$-lepton decay rate into nonstrange hadrons $H_{S = 0}$
that is given by
\be
\label{rate}
R_{\tau S=0}=\frac{\Gamma(\tau \rightarrow H_{S=0} \nu)}
{\Gamma(\tau \rightarrow l \bar{\nu} \nu )} \nn \\
= N_c |V_{ud}|^2 S_{EW}(1+\delta_P + \delta_{EW} + \delta_{NP})
\ee
where $N_c=3$ is the number of colors.
The first term in eq.~(\ref{rate}) is the parton model
result, the second term
$\delta_P$ represents perturbative QCD
effects. $V_{ud}$ is the flavor mixing matrix element~\cite{PDG}.
The factor $S_{EW}$ is an electroweak correction
term~\cite{ewcorr1} and $\delta_{EW} = 0.001$
is an additive electroweak correction~\cite{ewcorr2}.
The nonperturbative corrections are rather small
and consistent with zero;
$\delta_{NP} = -0.003 \pm 0.003$ (see e.g. \cite{NarPic88}).
For numerical estimates the
factorization approximation for four-quark condensates is essential:
it has been studied within $x$ space sum rules and found to be under
control~\cite{factor}.
Note that recently the problem of duality violation
for two-point correlators has been discussed \cite{chib,bigi}.
However, no established quantitative estimates
of that violation are available yet.
This problem can affect the numerical value of the coupling extracted
from the analysis because of the numerical change of the quantity
$\delta_{P}$ extracted from eq.~(\ref{rate}).
However, no established quantitative estimates
of that violation are available yet.
We concentrate on
the perturbative part of the decay
rate and numerical uncertainties related to
the renormalization scheme
freedom of perturbation theory.
In this respect new possible corrections do not qualitatively affect
our analysis. The corrections due to duality violation
are of a new nature and they can be added independently to
eq.~(\ref{rate}). They would only change the input
numerical value for the $\delta_{P}$ within our approach.

The value for the decay rate
$R_{\tau S=0}$ has been measured by the ALEPH~\cite{exp1} and OPAL~\cite{exp2}
collaborations with results very close to each other.
Using (for definiteness only) the ALEPH data
$R_{\tau S=0 }^{exp}=3.492 \pm 0.016$
one obtains from eq.~(\ref{rate}) a numerical value for the
main experimental parameter of the discussed observable
\be
\label{expdec0}
\delta_{P}^{\rm exp}=0.203\pm0.007 \ .
\ee
Eq.~(\ref{expdec0}) is the main experimental
input that is used in the analysis below.

Theoretically one starts with the determination of the
differential decay rate of the $\tau$ lepton into
a hadronic system  $H(s)$ with a total squared energy $s$
\[
\frac{d\sigma(\tau\to \nu H(s))}{ds}\sim \left(1-\frac{s}{M_\tau^2}\right)^2
\left(1+\frac{2s}{M_\tau^2}\right)\rho(s)
\]
that is given by the hadronic spectral density $\rho(s)$
defined through the correlator of weak currents. For the
$(ud)$ current $j_{\mu}^W(x) = \bar{u}\gamma_{\mu}(1-\gamma_5) d$
one finds in massless approximation for the light quarks
\be
\label{corr}
i\int\! \langle Tj_{\mu}^W(x)j_{\nu}^{W+}(0) \rangle e^{iqx}dx
=(q_\mu q_\nu - q^2 g_{\mu\nu})\Pi^{\rm had}(q^2),
\quad \Pi^{\rm had}(q^2)=\int \frac{\rho(s)ds}{s-q^2}
\ee
with $\rho(s)\sim {\rm Im}~\Pi^{\rm had}(s+i0)$, $s=q^2$.

Integrating the function $\Pi^{\rm had}(z)$
over a contour in the complex $q^2$ plane beyond the physical cut
$s>0$ one finds that
for particular weight functions some
integrals of the hadronic spectral density $\rho(s)$
can be reliably computed theoretically~\cite{cont1,cont2}.
Indeed, due to Cauchy theorem one gets
\[
\oint_C \Pi(z)dz = \int_{\rm cut} \rho(s) ds\ .
\]
Using the approximation
$\Pi^{\rm had}(z)|_{z\in C} \approx \Pi^{\rm th}(z)|_{z\in C}$
which is well justified sufficiently far from the physical cut
one obtains
\[
\oint_C \Pi^{\rm had}(z)dz = \int_{\rm cut} \rho(s) ds
= \oint_C \Pi^{\rm th}(z)dz
\]
i.e. the integral over the hadronic spectrum can theoretically be
evaluated. In practice, $\Pi^{\rm th}(z)$ is computed within OPE, i.e.
the function $\Pi^{\rm th}(Q^2)$ with $Q^2=-q^2$
is calculable in pQCD far from the physical cut
as a series in the running coupling constant $\al_s(Q^2)$
with power corrections. Further improvements on theory side can be
made -- for instance, instanton-induced contributions can be added.
The lattice approximation for the evaluation of
the correlator $\Pi^{\rm th}(Q^2)$ beyond perturbation theory can also be
used~\cite{lattice}.

The total decay rate of the $\tau$ lepton
written in the form of an integral along the cut
\[
R_{\tau S=0}=\frac{\Gamma(\tau \rightarrow H_{S=0} \nu)}
{\Gamma(\tau \rightarrow l \bar{\nu} \nu )}
\sim \int_{\rm cut}\left(1-\frac{s}{M_\tau^2}\right)^2
\left(1+\frac{2s}{M_\tau^2}\right)\rho(s)ds
\]
is precisely the quantity that one can reliably
compute in pQCD.

The basic object of the theoretical calculation is
Adler's $D$-function which is given by the
representation
\be
D(Q^2)=-Q^2\frac{d}{dQ^2}\Pi^{\rm had}(Q^2)\ .
\ee
In the $\overline{\rm MS}$-scheme the perturbative expansion for the
$D$-function is given by
\be
\label{spect0}
D(Q^2)=1+\frac{\al_s(Q)}{\pi}
+k_1 \left(\frac{\al_s(Q)}{\pi}\right)^2
+k_2 \left(\frac{\al_s(Q)}{\pi}\right)^3
+ k_3 \left(\frac{\al_s(Q)}{\pi}\right)^4+O(\al_s(Q)^5)
\ee
with (see e.g.~\cite{phys_report})
\be
  \label{ks}
k_1=\frac{299}{24} - 9\zeta(3), \quad
k_2=\frac{58057}{288} - \frac{779}{4}\zeta(3) +
\frac{75}{2}\zeta(5)\, .
\ee
The notation $a_s(Q)=\al_s(Q)/\pi$
for the standard $\MSsch$-coupling constant
normalized at the scale $\mu=Q$ is used.
Numerically one finds
\be
\label{spect}
D(Q^2)=1+a_s(Q)+1.6398 a_s(Q)^2 + 6.3710 a_s(Q)^3
+ k_3a_s(Q)^4 +O(a_s^5(Q))\, .
\ee
The coefficient $k_3$ is known only partly~\cite{Baikov:2001aa}.
The particular numerical value of $k_3\sim 25$ is
obtained on the basis of geometric series approximation for
the series (\ref{spect}) and is often used in the
literature~\cite{DP,kataev,Elias}.
Below I keep this coefficient for illustrative
purposes to see the potential influence of this term on
the numerical value of the coupling constant
extracted from $\tau$-data.

In the $\MSsch$-scheme
the perturbative correction $\delta_{P}$
is given by the perturbation theory expansion
\be
\label{taumssch}
\delta_P^{th} = a_s + 5.2023 a_s^2
+26.366a_s^3 +(78.003+k_3)a_s^4 + O(a_s^5)
\ee
where the $\MSsch$-scheme coupling constant
$\al_s=\pi a_s$ is taken at the scale of the $\tau$-lepton
mass $\mu=M_\tau = 1.777~{\rm GeV}$.
Usually one extracts a numerical value for $\al_s(M_\tau)$
by treating
the first three terms of the expression in eq.~(\ref{taumssch})
as an exact function -- the cubic polynomial, i.e. one solves
the equation
\ba
\label{taumsschEx}
a_s + 5.2023 a_s^2
+26.366a_s^3=\delta_P^{\rm exp}\, .
\ea
The solution reads
\be
\label{dirres}
\pi a_s^{st}(M_\tau)\equiv \al_s^{st}(M_\tau)= 0.3404\pm 0.0073_{exp} \, .
\ee
This is a standard method.
The quoted error is due to the error in the
input value of $\delta_P^{exp}$.
It is rather difficult to estimate the theoretical uncertainty of the
procedure itself. The main problem is to estimate the quality
of the approximation for the (asymptotic)
series in eq.~(\ref{taumssch}) given by
the cubic polynomial in eq.~(\ref{taumsschEx}).

As a criterion of the quality of the approximation
one can use the pattern of convergence of the series
(\ref{taumssch}) which is
\be
\delta_P^{\rm exp}=0.203=0.108+0.061+0.034+\ldots
\ee
One sees that the corrections provide a
100\% change of the leading term.
Another criterion is the order-by-order behavior of the extracted
numerical value for the coupling constant. In consecutive orders
of perturbation theory (LO -- leading order, NLO -- next-to-leading
order, NNLO -- next-next-to-leading order)
one has
\be
\al_s^{st}(M_\tau)_{LO}=0.6377,\quad
\al_s^{st}(M_\tau)_{NLO}=0.3882,\quad
\al_s^{st}(M_\tau)_{NNLO}=0.3404\, .
\ee
One obtains a series for the numerical value of the
coupling constant
\ba
\al_s^{st}(M_\tau)_{NNLO}&=&0.6377-0.2495-0.0478-\ldots
\ea
Up to the next-to-next-to-leading order
result (NNLO) we can take a half of the last term
as an estimate of the theoretical
uncertainty.
No rigorous justification can be given for such an assumption
about the accuracy of the approximation without knowledge
of the structure of the whole series: it is taken for definiteness.
The theoretical uncertainty obtained in
such a way --
$\Delta \al_s^{st}(M_\tau)_{th}=0.0239$ --
is much larger than the experimental uncertainty
given in eq.~(\ref{dirres}).
This is a challenge for the theory: the accuracy of
theoretical formulas cannot
compete with experimental precision at present.
Assuming this theoretical uncertainty we have
\be
\label{finalST}
\al_s^{st}(M_\tau)_{NNLO}=0.3404\pm 0.0239_{th}\pm 0.0073_{exp}.
\ee
Theory dominates the error even if the estimate for
its precision
$\pm 0.0239_{th}$ is not reliable (heuristic
and only indicative).
Thus the straightforward analysis in the $\MSsch$-scheme
is not stable numerically and the naive estimate of the theoretical
uncertainty is large.

To highlight the essence of the problem let me choose a different
coupling constant as an expansion parameter that is obtained by the
simple RG transformation -- the change of scale
of the coupling along the RG trajectory
$M_\tau\to 1~{\rm GeV}$.
It is often called scale transformation which is still a subgroup of
the full renormalization group.
In terms of $a_s(1~{\rm GeV})$ one finds a series
$a_s(1) +  2.615 a_s(1)^2
+1.54 a_s(1)^3=\delta_P^{\rm exp}$.
The solution for the coupling constant is $\al_s(1~{\rm GeV})=0.453$.
The convergence pattern for the correction
$\delta_P^{\rm exp}$ is
$\delta_P^{\rm exp}=0.203 = 0.144 + 0.054 + 0.005$
and for the numerical value of the coupling constant
$\al_s(1\GeV)= 0.453=0.638-0.177-0.008$.
Should one conclude that now the accuracy is much better?
What would be an invariant criterion for the precision
of theoretical predictions obtained from perturbation theory, i.e. finite number
of terms of asymptotic series?
Thus, one sees that the renormalization scheme dependence can
strongly obscure the heuristic evaluation of the accuracy of
theoretical formulae in the absence of any information on the
structure of the whole perturbation theory series.

The use of the $\MSsch$-scheme is not obligatory for
physical applications.
The $\MSsch$-scheme has a history of success for
massless calculations where its results look natural and
the corrections are usually small. This is not the strict rule, however,
and there are cases (like gluonic correlators \cite{glu})
where corrections dramatically depend on the quantum numbers
of the operators.
In fact, the $\MSsch$-scheme is rather artificial.
It is simply defined by convention
(let us be remindful of
the evolution from the MS-scheme to the $\MSsch$-scheme which had
its origin only in technical convenience \cite{buras}). From
technical point of view,
in practical calculations of massless diagrams of the propagator type,
another scheme -- the $G$-scheme --
is the most natural one \cite{Gsch}. It normalizes the basic quantity
of the whole calculation within
integration-by-parts technique -- one loop massless
scalar diagram -- to unity \cite{intbyparts}.
$\beta$-functions coincide in both schemes.
It could have well happened that the $G$-scheme would be historically
adopted as the reference scheme
because corrections in this scheme are typically smaller than that in
the $\MSsch$-scheme.
However, for the tau system the direct (standard)
analysis in the $G$-scheme fails.

Note that strictly speaking any scheme is suitable for
a given perturbative calculation.
However, it can lead to unusual (or even unacceptable) results
in a numerical analysis.
The only criterion for the choice of scheme at present is the
heuristic requirement of fast explicit convergence: the terms of the
series should decrease. Clearly this is a rather unreliable criterion.
It does not provide strict quantitative constraints necessary for the
level of precision usually claimed for the $\tau$-system analysis.

One can however work within a different paradigm
that is independent
of the scheme in what the actual calculation has been performed.
One just extracts a scale that any observable generates due to
dimensional transmutation in perturbation theory
and which is its internal scale.
It is natural for a numerical analysis
to determine this scale first.
For comparison with other channels one can then transform
the result into a $\MSsch$-scheme
or any other reference scheme
using the renormalization group invariance.
This last step is done only for comparison
with other experiments (or just for convenience;
the system itself can be well described in its internal scheme
without any reference to the $\MSsch$-scheme).

The running of the coupling is one of the central features of QCD and
very important for numerical analysis.
A parameterization of the trajectory can be done at infinite $Q$ by
using dimensional scale of QCD.
The renormalization group equation
\be \label{RGE}
\mu^2 \frac{d}{d \mu^2} a(\mu^2) = \beta(a(\mu^2))\, ,
\quad a=\frac{\al}{\pi}
\ee
is solved by the integral
\be
\label{LQCD}
\ln \left(\frac{\mu^2}{\Lambda^2}\right) =
   \Phi(a(\mu^2)) + \int_0^{a(\mu^2)}
\left( \frac{1}{\beta(\xi)} - \frac{1}{\beta_2(\xi)} \right)d\xi
\ee
where the indefinite integral $\Phi(a)$ is normalized
as follows
\be
\label{intb2}
\Phi(a) = \int^{a}  \frac{1}{\beta_2(\xi)} d \xi
= \frac{1}{a \beta_0}
       + \frac{\beta_1}{\beta_0^2}
\ln \left( \frac{a \beta_0^2}{\beta_0 + a \beta_1} \right).
\ee
Here
$\beta_2(a)$ and $\beta(a)$ denote the second order
and full $\beta$ function, or as many terms as are available,
given by
\be
\beta_2(a)= -a^2(\beta_0 + a \beta_1), \quad
\beta(a)= -a^2(\beta_0 + \beta_1 a + \beta_2 a^2
+ \beta_3a^3) +O(a^6)\, ,
\ee
$a$ is a generic coupling constant.
The four-loop $\beta$-function coefficient $\beta_3$ is now known in
the $\MSsch$-scheme~\cite{beta4}
\be
\beta_3=\frac{140599}{4608} + \frac{445}{32}\zeta(3)=47.228\ldots
\ee
The adjustment of the integration constant in eq.~(\ref{LQCD})
defines the standard QCD scale $\Lambda_s$: the asymptotic expansion
of the coupling constant at large momenta $Q^2\to \infty$ reads
\be
\label{Lser}
a(Q^2)= \frac{1}{\beta_0 L}\left(1 - \frac{\beta_1}{\beta_0^2}
\frac{\ln(L)}{L} \right)+ O \left(\frac{1}{L^3} \right)\, ,
\qquad L=\ln\left(\frac{Q^2}{\Lambda^2}\right) \, .
\ee
The solution (\ref{LQCD}) of the renormalization group equation (\ref{RGE})
describes the evolution trajectory of the coupling constant.
The evolution trajectory of the coupling constant
given by the solution (\ref{LQCD}) of the renormalization group
equation (\ref{RGE}) is parametrized by the scale parameter
$\Lambda$ and the coefficients of the $\beta$ function
$\beta_i$ with $i>2$ (see e.g. \cite{stevenson}). The evolution is
invariant under the renormalization group transformation
\be
\label{kappa}
a  \rightarrow a(1 + \kappa_1 a + \kappa_2 a^2 + \kappa_3 a^3 + \dots )
\ee
with the simultaneous change
\be
\label{transform}
\Lambda^2 \rightarrow  \Lambda^2 e^{-\kappa_1/\beta_0}\, ,
\ee
$\beta_{0,1}$ left invariant and
\ba
\beta_2 &\rightarrow&\beta_2-\kappa_1^2\beta_0
+\kappa_2 \beta_0 - \kappa_1 \beta_1 \nn \\
\beta_3 &\rightarrow&\beta_3+4 \kappa_1^3 \beta_0
+ 2 \kappa_3 \beta_0 + \kappa_1^2
\beta_1 - 2 \kappa_1 (3 \kappa_2 \beta_0 + \beta_2 )  .\nn
\ea
If this transformation was considered to be exact
and the exact $\beta$-function
corresponding to the new charge was used then it would be just a
change of variable in a differential equation (\ref{RGE}) or
the exact reparametrization of the trajectory (\ref{LQCD})
and hence would lead to identical results.
However, the renormalization group invariance of eq.~(\ref{LQCD})
is violated in higher orders
of the coupling constant because we consistently omit higher orders
in the perturbation theory expressions for
the $\beta$-functions.
This is the point where the finite-order perturbation
theory approximation for the respective $\beta$-functions
is made.
This is the source for different numerical outputs of analyses in
different schemes.

Then the paradigm is to use the optimal charge and
extract the coupling using exact RG technique.
One introduces an effective charge $a_\tau$ through the
relation
$\delta_P^{th} = a_\tau\equiv \al_\tau/
\pi$~\cite{prl,grunberg,krasK,dhar,brodsky}
and extracts the parameter $\Lambda_\tau$ which
is associated with $a_\tau$ through eq.~(\ref{LQCD}).
This is just the internal scale associated with
the physical observable $R_\tau$.
The effective $\beta$-function is given by the expression
\be
\beta_\tau = -a_\tau^2(\beta_{\tau 0} + \beta_{\tau 1} a_\tau +
\beta_{\tau 2} a_\tau^2 + \beta_{\tau 3} a_\tau^3 + \dots)
\ee
with
$\beta_{\tau 0} = \beta_0$, $\beta_{\tau 1 } = \beta_1$, and
\be
\label{effbeta}
\beta_{\tau 2 }= -12.3204\, ,
\quad
\beta_{\tau 3 }= -182.719 + \frac{9}{2} k_3 \, .
\ee
The extraction of the numerical value for the internal scale
$\Lambda_\tau$ is done
from equation (\ref{LQCD}) with
$a_\tau(M_\tau) = \delta_P^{\rm exp}$.
The coefficient $\beta_{\tau 3}$ does not enter the analysis.
The parameter $\Lambda_s\equiv \Lambda_{\MSsch}$ is found
according to eq.~(\ref{transform}).
The $\MSsch$ coupling at $\mu = M_\tau$
is obtained by solving eq.~(\ref{LQCD})
for $a_s(M_\tau)$ with regard to $\ln(M_\tau^2 / \Lambda_s^2)$
which is known if $\Lambda_s$ is obtained; the
$\beta$-function is taken in the $\MSsch$-scheme.
For consistency reasons we only use the $\MSsch$-scheme
$\beta$-function
to three-loop order since the effective $\beta$-function $\beta_\tau$
is only known up to the second order, cf. eq. (\ref{effbeta}).
A $N^3LO$ analysis is possible only if a definite value
is chosen for $k_3$.
Some estimates are given below.

The procedure is based on
renormalization group invariance and one can start from the
expression for the decay rate obtained in any scheme.
The only perturbative objects present are the $\beta$-functions:
it is the converge of perturbation theory for $\beta$-functions that determines
the accuracy.
It also highlights the limit of precision within this procedure:
the expansion for $\beta_\tau$ is believed to be asymptotic
as any expansion in perturbation theory.
The asymptotic expansion provides only limited accuracy
for any given numerical value of the expansion parameter
which cannot be further improved by including higher order terms.
The expansion used is presumably rather
close to its asymptotic limit
\be
\label{betatau}
\beta_\tau(a_\tau)=- a_\tau^2 \left(\frac{9}{4} + 4 a_\tau
- 12.3204 a_\tau^2
+ a_\tau^3 \left(-182.719+\frac{9}{2} k_3 \right) \right)
+O(a_\tau^6)
\ee
with $a_\tau\sim 0.2$ at the scale $M_\tau$.
The convergence of the
series depends crucially on the numerical value of $k_3$.
If $k_3$ had a value where
the asymptotic growth starts at third order
then further improvement of the accuracy within
finite-order perturbation theory
is impossible.

At every order of the analysis we use the whole information of
the perturbation theory
calculation.
Especially, the appropriate coefficient of the $\beta_\tau$-function
is present.
In the standard method the coefficient $\beta_2$
enters only at order $O(\al_s^4)$ of the
$\tau$-lepton decay rate expansion.
The described procedure can be called
the renormalization scheme invariant extraction
method (RSI) that means that it is based on
renormalization scheme invariance. It does not mean that
renormalization scheme uncertainty is completely eliminated but
that it put under control in terms of explicit convergence of the
effective $\beta$ function.
It is also clear since $\al_s$ itself is not
a physical
object and is not renormalization scheme invariant.
In this respect we extract the noninvariant
parameter $\al_s$ using invariance of the physics in order to perform
the numerical analysis in the most suitable scheme.
The output of the
analysis is transformed into a numerical value for $\al_s$
according to the renormalization group transformation rules
that are treated exactly for the given order of the effective
$\beta$ function.
For the coupling constant in the $\MSsch$-scheme in NNLO
it gives
\ba
\label{finFO}
\al_{s}^{RSI}(M_\tau) &=& 0.3184 \pm 0.0060_{exp}
\ea
which is smaller than the corresponding value obtained within
the standard procedure eq.~(\ref{dirres}).
The key question is how to estimate the quality of this result?
The parameter which is really extracted
in consecutive orders of perturbation theory
within the method is
the scale $\Lambda_\tau$. Because of the relation
\be
\Lambda_s=\Lambda_\tau e^{-5.20232/2\beta_0}=0.3147\Lambda_\tau
\ee
one can look at $\Lambda_s$ directly.
One finds
\be
\Lambda_s|_{LO}=595~{\rm MeV},\quad
\Lambda_s|_{NLO}=288~{\rm MeV},\quad
\Lambda_s|_{NNLO}=349~{\rm MeV}
\ee
or, representing the NNLO result as a formal series,
\be
\label{lamser}
\Lambda_s|_{NNLO}=595-307+61-\ldots~{\rm MeV}.
\ee
Note that at leading order the scales (as well as charges)
are equal in all schemes. Therefore the leading order result
($\Lambda_s|_{LO}=595~{\rm MeV}$) is not representative,
only indicative.
Assuming the uncertainty of $\Lambda_s$
to be given by the half of the last term of the series (\ref{lamser})
one has $\Lambda_s=349\pm 31~{\rm MeV}$
which leads to the numerical value for the $\MSsch$-scheme coupling
constant
\be
\al_s=0.3184^{-0.0157}_{+0.0160} \, .
\ee
This result is obtained from eq.~(\ref{LQCD}) with three-loop
$\beta$-function. Taking the average we find
\be
\label{theoryNNLO}
\al_s=0.3184 \pm 0.0159\, .
\ee
This is better than the theoretical error of the standard result
eq.~(\ref{finalST}).
Still the theoretical error should be considered as a guess
rather than a well-justified estimate of the uncertainty.
It is affected by higher order terms, e.g.
by the $k_3$ contribution.
Clearly the estimate $k_3=25$ is rather speculative.
Therefore it is more instructive to determine the range of $k_3$
which is safe for explicit convergence
of perturbation theory. If the actual value of $k_3$ will be
discovered in this range
then perturbation theory is still valid
and will give better accuracy in NNNLO.
If not, the asymptotic growth of perturbation theory series is
already reached and its accuracy cannot be improved.

We require that the last term should be equal to the half of
the previous one. In the standard way (eq.~\ref{taumssch}) we have
\be
|(78+k_3)a_s|<\frac{1}{2}\, 26.36 \approx 13
\ee
which for $a_s=0.1$ gives
\be
\label{standrest}
-208<k_3^{st}<52 \, .
\ee
In the RSI way (eq.~\ref{betatau}) we have
\be
|(-182+\frac{9}{2}k_3)a_\tau|<\frac{1}{2}\, 12.32 \approx 6
\ee
which for $a_\tau=0.2$ gives
\be
33.8<k_3^{\tau}<47.1 \, .
\ee
This range is much narrower
than that in eq.~(\ref{standrest}).
The effective scheme method is much more sensitive
to the structure of the series as can be seen from eq.~(\ref{betatau}).
The
actual precision depends on the actual value chosen for $k_3$
and it is rather premature to speculate about
numbers.

Still we show the worst result
(in the optimistic scenario that $k_3$ lies in the safe
range)
that can be expected within the RSI approach.
In the RSI approach with $k_3=47$ we find the scale parameter in NNNLO
\be
\Lambda_s|_{NNNLO}=334~{\rm MeV} \, .
\ee
With $k_3=34$ one has
\be
\Lambda_s|_{NNNLO}=367~{\rm MeV}  \, .
\ee
Taking the average we have
\be
\Lambda_s=350\pm 17~{\rm MeV}
\ee
which is the best possible
estimate if we require that the perturbation theory series for
the $\beta_\tau$-function still
converges (according to our quantitative criterion of convergence).
That results in the numerical value for the $\MSsch$-scheme coupling
constant found with four-loop $\beta$-function from eq.~(\ref{LQCD})
\be
0.3133< \al_s< 0.3314 \, .
\ee
Therefore our conservative estimate of the theoretical error
in the optimistic scenario
for the convergence of perturbation theory series in NNNLO
reads
\be
\label{optNNNLO}
\al_s=0.322\pm 0.009\, .
\ee
One should keep in mind that
the theoretical uncertainty still depends on the criteria chosen.
This particular estimate comes from the requirement of the explicit
convergence of the effective beta function.

The value of the coupling constant can be run to
the scale $M_Z = 91.187~{\rm GeV}$ with the RG
For the standard method one finds
\be
\label{dir0}
\al_s^{st}(M_Z)=0.1210 \pm 0.0008_{exp} \pm 0.0006_c \pm 0.0001_b
\ee
where the subscript $exp$ denotes the error
originating from $\delta_P^{exp}$.
The errors with subscripts
$c,b$ arise from the uncertainty of the numerical values of
the charm and bottom quark
masses that enter the evolution analysis.
The running to this reference scale
is done with the four-loop $\beta$-function in
the $\MSsch$-scheme \cite{beta4}
and three-loop matching conditions
at the heavy quark (charm and bottom)
thresholds \cite{matching}.
The threshold parameters related to heavy quark masses
are
$\mu_c=\bar{m}_c(\mu_c)=(1.35\pm 0.15)~{\rm GeV}$
and
$\mu_b=\bar{m}_b(\mu_b)=(4.21\pm 0.11)~{\rm GeV}$ (e.g.~\cite{bbmass})
where $\bar{m}_q(\mu)$ is the running mass of the heavy quark in
the $\MSsch$-scheme.

The running of $\al_s^{RSI}(M_\tau)$
from eq.~(\ref{theoryNNLO}) gives
\be
\label{rgi0}
\al_{s}^{RSI}(M_Z) = 0.1184 \pm 0.00074_{exp}
\pm 0.00053_c \pm 0.00005_b
\ee
The theoretical uncertainty comes mainly from
the truncation of the perturbation theory series.
Taking the result of the NNLO analysis eq.~(\ref{theoryNNLO})
one finds
\be
\Delta \al_s^{RSI}(M_Z)_{th}=0.0019
\ee
In the most optimistic scenario with the NNNLO
analysis eq.~(\ref{optNNNLO}) one has
\be
\al_s^{RSI}(M_Z)_{N^3LO}=0.119\pm0.001\, .
\ee
Note that for the Cabibbo suppressed decays that can also be
analyzed within the perturbation
theory~\cite{msNPBPP,others2,others3}
the approach looks quite
natural as there are additional functions to be studied:
two functions related to $m_s^2$ perturbative corrections.
Then three coefficient functions should be analyzed together
that allows one to factor out the renormalization scheme dependence
to large extent~\cite{krajms}.

Now I briefly comment on the extraction of
the strong coupling constant within resummed
perturbation theory. There are basically
two possibilities to resum perturbation theory
series that have recently been analyzed in some
details~\cite{antiren}.
One is to
extrapolate the running of the coupling into the infrared region using a
natural cutoff provided by analytic continuation (so called $\pi^2$
terms~\cite{pi2,shirk}). This approach relies on the pattern of continuation
of the coupling constant to the infrared region.
Another approach is based on the integration along the contour in
Euclidean domain, it bypass the potentially nonperturbative region
along safe areas in the complex momentum plane and, therefore, is
quite perturbative in nature~\cite{Pivtau}.
Numerically the results of these two
approaches are different. In fact, they are also nonequivalent
mathematically and the difference can be explicitly
calculated: it happens to be nonexpandable in $al_s$ that means that
it is not noticeable within perturbation theory.
As for the numerical results one fits the theoretical expression for the
decay rate in the contour improved approach
to the experimental
result $\delta_{P}^{\rm exp}$ eq.~(\ref{expdec0}) and find
\be
\label{CI}
\al_s^{CI}(M_\tau) = 0.343 \pm 0.009_{exp}
\ee
within the renormalization scheme invariant extraction
method described above i.e. with the introduction of
the effective charge first~\cite{resana}.
This value differs from the finite-order perturbation theory result
eq.~(\ref{finFO}) which is expected.

To conclude, the procedure of extracting the numerical
value of the strong coupling constant from $\tau$-data
is briefly described with main emphasis on the particular features
that are related to
the renormalization scheme invariance of the theory.

The present work is supported in part by the Russian Fund for
Basic Research under contract 02-01-00601 and by INTAS grant.

\end{document}